\documentclass[letterpaper, 10pt, conference]{ieeeconf}
\IEEEoverridecommandlockouts 

\usepackage{tabularx}
\usepackage[utf8]{inputenc}
\usepackage{amsmath,amssymb,amsfonts}
\usepackage{textgreek}
\usepackage{graphicx}
\usepackage{color,bm}
\usepackage{wasysym}
\usepackage{tikz}
\usepackage{cite}
\usepackage{array,multirow}
\usepackage{float}
\usepackage{balance}
\usepackage[hidelinks]{hyperref}
\usepackage{mathtools, nccmath}
\DeclareGraphicsExtensions{.tif}
\usepackage[linesnumbered,ruled,vlined]{algorithm2e}
\usepackage{amsmath} 
\usepackage{algorithmic,algorithm2e}
\usepackage{booktabs}

\usepackage{geometry}
\usepackage{url}

\newcount\Comments  
\newcount\Revision 
\definecolor{darkblue}{rgb}{0,0,0.75}
\newcommand{\kibitzrev}[2]{\ifnum\Revision=0\textcolor{#1}{#2}\else\textcolor{black}{#2}\fi}

\title{\LARGE \textbf{Real-time Learning of Driving Gap Preference for Personalized Adaptive Cruise Control}}

\author{Zhouqiao Zhao,
        Xishun Liao,
        Amr Abdelraouf,
        Kyungtae Han,
        Rohit Gupta, \\
        Matthew J. Barth
        and Guoyuan Wu
\thanks{Z. Zhao, X. Liao, M. Barth, and G. Wu are with College of Engineering, University of California, Riverside, 1084 Columbia Avenue, Riverside, CA 92507 (e-mail:
zzhao084@ucr.edu;
xliao016@ucr.edu;
barth@ece.ucr.edu;
gywu@cert.ucr.edu).}
\thanks{A. Abdelraouf, K. Han, and R. Gupta are with InfoTech Labs, Toyota Motor North America R\&D, 465 Bernardo Avenue, Mountain View, CA 94043 (e-mail:
amr.abdelraouf@toyota.com;
kyungtae.han@toyota.com;
rohit.gupta@toyota.com).}
}

\begin{document}
\maketitle
\begin{abstract}
Advanced Driver Assistance Systems (ADAS) are increasingly important in improving driving safety and comfort, with Adaptive Cruise Control (ACC) being one of the most widely used. However, pre-defined ACC settings may not always align with driver's preferences and habits, leading to discomfort and potential safety issues. Personalized ACC (P-ACC) has been proposed to address this problem, but most existing research uses historical driving data to imitate behaviors that conform to driver preferences, neglecting real-time driver feedback. To bridge this gap, we propose a cloud-vehicle collaborative P-ACC framework that incorporates driver feedback adaptation in real time. The framework is divided into offline and online parts. The offline component records the driver's naturalistic car-following trajectory and uses inverse reinforcement learning (IRL) to train the model on the cloud. In the online component, driver feedback is used to update the driving gap preference in real time. The model is then retrained on the cloud with driver's takeover trajectories, achieving incremental learning to better match driver's preference. Human-in-the-loop (HuiL) simulation experiments demonstrate that our proposed method significantly reduces driver intervention in automatic control systems by up to 62.8\%. By incorporating real-time driver feedback, our approach enhances the comfort and safety of P-ACC, providing a personalized and adaptable driving experience.
\end{abstract}
 
\section{Introduction}
The last decade has witnessed the rapid emergence and booming development of vehicle automation and Advanced Driver Assistance Systems. As technology continues to advance, vehicles are becoming more intelligent, with the ability to perform tasks that were once solely the responsibility of the driver. ADAS systems are becoming more sophisticated, providing drivers with a range of features that improve safety and convenience, such as adaptive cruise control (ACC), lane departure warning (LDW), and automated emergency braking (AEB)  \cite{TSS, VWACC, FordACC}. In addition to enhancing the driving experience, these technologies also have the potential to reduce accidents, injuries, and fatalities on the road \cite{ziebinski2017review}.

While Advanced Driver Assistance Systems (ADAS) and vehicle automation have the potential to improve road safety and driving comfort, a lack of personalization can lead to various issues. Without personalized settings, drivers may experience discomfort, reduced trust in the automation system, decreased usage, and increased risk of accidents due to unintended operations. Additionally, individuals have different driving habits and preferences, so a one-size-fits-all approach may not work for everyone \cite{hasenjager2019survey}. Personalization of ADAS and vehicle automation can help to address these issues, tailoring the system's settings to meet the specific needs and preferences of each driver. This can not only improve driving comfort and trust in the automation system but also increase usage and reduce the likelihood of accidents caused by user error.

In this work, we propose a Personalized Adaptive Cruise Control (P-ACC) system that learns both from the naturalistic car-following behaviors of individual drivers (e.g., offline learning) and from the drivers' real-time feedback (i.e., online learning). The offline personalization is achieved by the Inverse Reinforcement Learning (IRL) algorithm, which infers the reward function of a target agent given the demonstration trajectories. The recovered reward function is a representation of the driver's driving style, which is then transferred into a driving gap preference table (DGPT) and sent to the controller as the reference. This DGPT is utilized to control the vehicle while ACC is engaged. Instead of simply cloning the driver's behavior from data, the recovered reward function helps explain the observed demonstrations and the driver's task-specific preferences. The online personalization is achieved by a heuristic algorithm that adaptively adjusts the control reference DGPT in real-time based on the driver's feedback as expressed through ACC overrides, such as applying the accelerator or brake pedals. Compared to the naturalistic car-following data that is used for training the IRL model, the online feedback data usually has a very short time period and distribute sparsely in the time domain. Therefore, additional processes are required to use this data to maintain the DGPT. Additionally, the driver's feedback data is used to update the offline personalization module as an incremental learning scheme after completing a trip. By combining both offline and online learning, our proposed P-ACC system can provide a personalized driving experience for each driver, resulting in enhanced driving comfort and safety.

Compared to the existing literature that studied the personalization of ACC systems, we make the following contributions in this work:
\begin{itemize}
    \item We propose a novel Cloud-Vehicle P-ACC framework. This framework enables the P-ACC system to learn from a large number of drivers' data and adapt to their unique driving styles. By leveraging cloud-based computation and communication, our framework is able to process a large amount of driving data, allowing for more accurate personalization of the P-ACC system.
    \item We introduce an offline-online scheme for driving personalization. The offline scheme uses the Inverse Reinforcement Learning (IRL) algorithm to recover the personalized driving style of each driver based on demonstration trajectories. The online scheme utilizes a heuristic algorithm to adaptively adjust the P-ACC system's behavior in real-time based on the driver's feedback. The combination of the offline IRL algorithm and the online heuristic algorithm ensures that the P-ACC system provides a personalized driving experience that aligns with the driver's preferences.
    \item We conduct a simulation experiment using the HuiL driving simulator to evaluate the effectiveness of our proposed P-ACC system. The experiment involves multiple drivers, and the results demonstrate the superior performance of our P-ACC system compared to existing ACC systems.
\end{itemize}

The remainder of this work is organized as follows: Section \ref{sec:review} reviews the latest literature in the related field. Section \ref{sec:problem} introduces the problem formulation. Section \ref{sec:method} elaborates on the proposed system. In Section \ref{sec:experiment}, we conduct numerical experiments on the naturalistic driving data and human-driving experiments on a game engine-based simulator to test the validity of the model. Finally, the study is concluded with some future directions in section \ref{sec:conclusion}.

\begin{figure*}
    \centering
    \includegraphics[width=\linewidth]{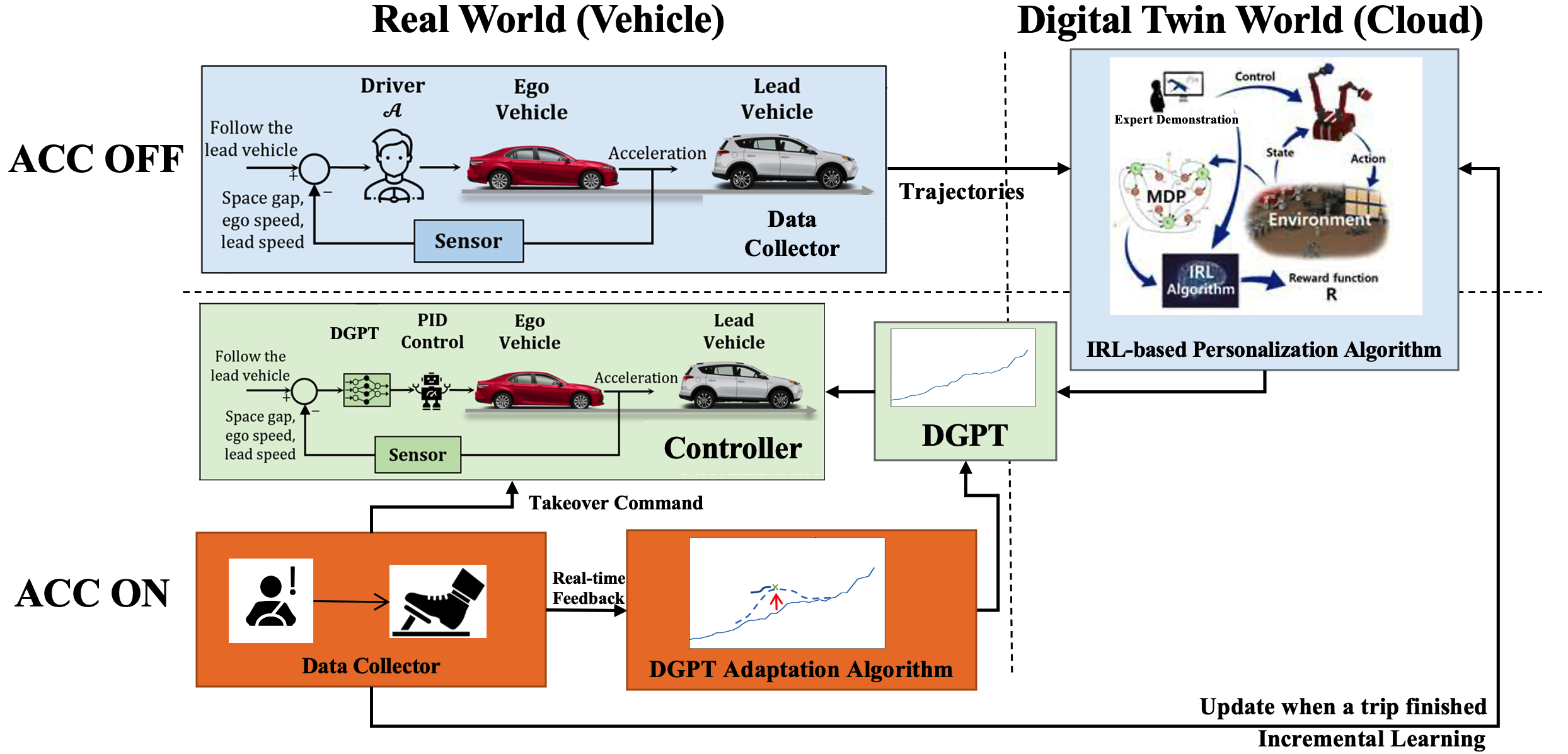}
    \caption{System architecture: Offline learning (blue blocks), Online learning (orange blocks), Personalized controller (green blocks).}
    \label{fig:architecture}
\end{figure*}

\section{Related Work} \label{sec:review}
\subsection{Car-following Model}
The foundation of the ACC system is based on the car-following model, which aims to replicate or optimize the behavior of a human driver in maintaining a safe and comfortable distance from the vehicle in front of them while driving. Existing literature on car-following modeling can be broadly classified into several categories, including Ordinary Differential Equation (ODE), Model Predictive Control (MPC), Inverse Reinforcement Learning (IRL), Gaussian Process Regression, and Sequential models. 

ODE-based policies aim to enable the ego vehicle to follow the movement of the preceding vehicle based on a set of predefined parameters. However, designing corresponding algorithms requires prior knowledge of the car-following system (e.g. vehicle dynamics), making them generic and difficult to personalize. Moreover, ODE and MPC policies lack expressivity, which makes it hard to capture the nuances of naturalistic human driving. Studies like \cite{Gipps1981TR}, \cite{Kesting2010PT}, and \cite{Newell2002TR} fall under this category. MPC, on the other hand, uses Model Predictive Control to optimize predefined objectives like safety, comfort, and fuel efficiency in a receding-horizon fashion \cite{Shengbo2011TCST}. Similar to ODE-based policies, designing MPC policies also requires prior knowledge of the car-following system, making them generic and difficult to personalize. IRL is another popular approach to learn personalized car-following behaviors. Researchers in \cite{Zhao2022ICRA} and \cite{Gao2018JARS} use IRL to learn the reward of car-following demonstration trajectories and implement the recovered reward using controllers. The IRL algorithms used in these studies can recover personalized car-following gap preferences based on different vehicle speed values, which can be used to design the downstream control logic for P-ACC systems. Gaussian Process Regression is a direct approach that looks into the data and learns from demonstration trajectories. Researchers in \cite{Wang2021ITSC} propose a Gaussian Process Regression algorithm for P-ACC, where both numerical and human-in-the-loop experiments verify the effectiveness of the proposed algorithm in reducing the interference frequency by the driver. Finally, since the decision-making process of human drivers depends on sequential state inputs, Recurrent Neural Network (RNN) and Long Short Term Memory (LSTM) have also been used to model car-following behaviors. Studies like \cite{Chong2011TRR} and \cite{Huang2018} fall under this category.

\subsection{Personalized Driving Behavior Modeling}
While driving behavior and preferences can be diverse among drivers, there is a growing demand to explore personalized driving behavior to enhance the safety and user experience of the current ACC system.

Driving style is widely adopted to modeling the personalized behavior in a high-level, namely the emotional and intentional level, as it can provide valuable insights into a driver's habits, preferences, and tendencies. Considering the driving style divergence among drivers, \cite{zhu2019typical} proposed a P-ACC with driving style identification and corresponding personalized speed-distance control. The driving style is characterized by fitting driving data of each individual driver into a Gaussian mixture model (GMM) and clustered by Kullback–Leibler (KL) divergence. Instead of only considering ego driver, personalized driving style is also depend on the environment. \cite{bao2021prediction} employed Conditional Variational Auto-Encoder to model a probabilistic distribution of the individual’s driving style considering surrounding vehicles, to facilitate the prediction for a driver's longitudinal acceleration and speed. 

Besides modeling a high-level driving style, Imitation Learning is another popular approach to model personalized driving behavior from demonstration. By observing the demonstration of the studied individual, IRL was implemented to recover the cost function \cite{liao2023driver}\cite{chen2020porf} for representing a driver's preference or a reward \cite{Zhao2022ICRA} for optimal policy. Similarly, Generative Adversarial Imitation Learning (GAIL) was used to learn the personalized car-following strategy only based on drivers’ demonstrations but without specifying the reward\cite{s20185034}. 

Moreover, researchers developed end-to-end approaches with integrating personalized behavior implicitly. To model the uncertainty of human behavior, a Gaussian Process Regression \cite{Wang2021ITSC} was adopted to learn the personalized longitudinal driving behavior model, which is a joint Gaussian distribution mapping from the driver’s perceived states to control outputs. \cite{huang2021personalized} utilized constraint Delaunay triangulation to identify a safe area and the fuzzy linguistic preference relation (FLPR) method to determine drivers' driving preferences. With taking the user's personalized objectives as input, this method achieved personalized trajectory planning and lane-change control, meeting users' diverse preferences while ensuring vehicle safety.

However, the mentioned driving behavior modeling can be affected by many factors, such as weather, road conditions, and the emotional state of the driver\cite{liao2022driver}. Existing literature only relies on historical data for personalized driving behavior modeling, which may not account for changes in external factors. Thus, to improve the flexibility and effectiveness of P-ACC, it is crucial to incorporate real-time data and driver feedback to dynamically adjust the car-following policy.

\section{Problem Formulation} \label{sec:problem}
\subsection{System Architecture}
As an extension of our previous study, this paper proposes a system architecture that is based on the vehicle-cloud framework  \cite{Zhao2022ICRA}, called ``Digital Twin''. This framework uploads drivers' naturalistic driving trajectory data and ACC feedback data to the cloud, where personalized models for different drivers are maintained. In this way, the cloud can effectively share the high computational demand of the training process. For similar driving behaviors, remote models can be reused to reduce the number of models. Correspondingly, data from similar driving behaviors can be used to fine-tune the same model for federated learning. Additionally, for the same or similar types of drivers facing different driving scenarios, such as different weather conditions, traffic conditions, or road conditions, the cloud can maintain different models for each scenario to achieve more precise personalization. For more detailed descriptions and implementations of Digital Twin, please refer to our previous article.

The proposed P-ACC system architecture, illustrated in Fig. \ref{fig:architecture}, is distinct from our previous study in which we only relied on modeling the demonstration car-following trajectory (blue blocks in Fig. \ref{fig:architecture}). In contrast, this paper introduces a novel approach to incorporate the driver's real-time feedback on the ACC system as a dynamic input to adjust the model  (orange blocks in Fig. \ref{fig:architecture}). Based on our literature review, such an approach has not been explored in prior research. The physical layer of the framework is divided into the real world (vehicle) and the digital twin world (cloud), while the implementation process is divided into two phases: ACC OFF and ACC ON.

During the ACC OFF phase, when the driver manually follows the lead vehicle, the system considers the trajectory as an expert demonstration and transmits it to the cloud along with environmental factors that could potentially impact driving behavior. On the cloud, the IRL algorithm assumes that the collected expert demonstration is near-optimal in terms of the Markov Decision Process (MDP) and infers the reward function that drives the driver's behavior. This reward function is then transferred to the DGPT as the control reference.

During the ACC ON phase, when the driver turns on the automatic following mode, the personalized driving model (i.e., DGPT) is downloaded locally. The DGPT is designed to describe the driver's preferred following distance at different speeds. The controller (green blocks in Fig. \ref{fig:architecture}) maintains the distance to the lead vehicle. However, due to differences between the  scenario of demonstration trajectories and the current driving scenario, as well as changes in the driver's driving habits or mood, the driver may not always be satisfied with the current automated control. Therefore, the driver can provide feedback to the system by pushing the accelerator to shorten the car-following gap or brake pedals to lengthen the gap, leading to a takeover of the vehicle. These takeover segments are used to adjust the DGPT in real-time, responding to the driver's behavior. When the current ACC ON trip is completed, these takeover segments are sent back to the cloud to fine-tune the IRL model, improving the personalization of the system.

\subsection{Assumptions and Specifications}
In this paper, we focus on modeling and controlling personalized car-following maneuvers based on states of the ego vehicle and its preceding vehicle. Although the real-world ACC needs to manage both car-following stage and free-flow stage, the ACC discussed in this paper only includes the car-following stage, i.e., we assume the leader of the ego vehicle is always present. Also, only the longitudinal movement is observed and controlled. The goal of this work is to design speed control strategy for the ego vehicle to satisfy the driver's preference. 

We assume that the personalized car-following behavior of a driver can be described using DGPT, which corresponds to the preferred following distance of drivers at different speeds. Therefore, we also describe the car-following dynamic model in a 2D space spanned by the speed $v$ and the distance $g$ to the preceding vehicle. We use a second-order approximation and discretize the space and speed, using the following equation:
\begin{equation}\label{eq:1}
    v[t+1]=v[t]+a[t] \cdot \Delta t+\sigma_{v}
\end{equation}
\begin{equation}\label{eq:2}
    g[t+1]=g[t]+\left(v_{f}[t]-v[t]\right) \cdot \Delta t+\frac{1}{2} \cdot a[t] \cdot \Delta t^{2}+\sigma_{g}
\end{equation}
As shown in Equation (\ref{eq:1}) and (\ref{eq:2}), we add Gaussian noises, $\sigma_v$ and $\sigma_g$, denoting imperfectness of the driver's observation and control.

We presume the driver's decision-making process is a MDP defined by a five-tuple $\{S,U,T,r,\gamma\}$, where $S$ is the state space spanned by $v$ and $g$; $U$ is the one-dimensional action space of all possible acceleration of the ego vehicle; $T$ is the transition probability determined based on Equation (\ref{eq:1}) and (\ref{eq:2}); $r$ is the reward function that represents the driver’s personalized car-following style; and $\gamma$ is the discount factor weighting the importance of the historical rewards; At each time step, the process in certain state $v$ and $g$, and the driver may choose any action $a$. The process responds at the next time step by moving into a new state $s'$ based on $T$. Notably, although the speed of the preceding vehicle, $v_f$, is considered in MDP, it can be observed while driving. It should be noted that we also assume the human driver is rational and his/her actions are optimizing a cumulative reward function formulated as follows:
\begin{equation}\label{eq:3}
    v(\xi)=\sum_{t=0}^{N} \gamma^{t} \cdot r_{t}(s)=\sum_{t=0}^{N} \gamma^{t} \cdot \boldsymbol{\alpha}^{T} \cdot \boldsymbol{\Phi}(s)
\end{equation}
where $N$ denotes the time horizon, and $\xi$ denotes the trajectory of the ego vehicle. As seen in Equation (\ref{eq:3}), the instantaneous reward $r_t(s)$ is assumed to be expressed in a span of the reward basis $\boldsymbol{\Phi}$, whose dimension equals the total number of features, and $\boldsymbol{\alpha}$ stands for a vector of weight defining the linear combination. Additionally, it is assumed that the collected trajectory can reflect the drivers' driving style and that drivers are comfortable with their own driving style.

\section{Methodology} \label{sec:method}
\begin{figure}
    \centering
    \includegraphics[width=\linewidth]{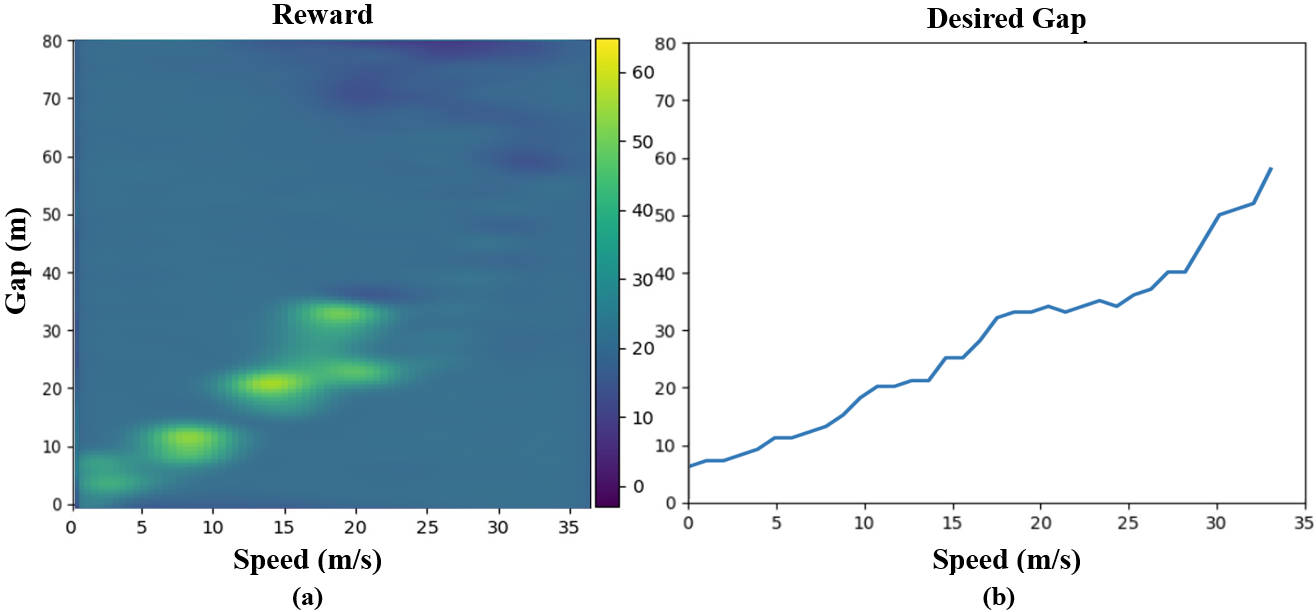}
    \caption{Personalized driving preference from IRL modeling: (a) recovered reward function from naturalistic driving data using IRL, (b) smoothed $speed$-$g_{desired}$ table.}
    \label{fig:reward_desired_gap}
\end{figure}

In this section, we present a detailed description of the proposed system, which includes modeling the driver's preference using the IRL algorithm, the online adaptation algorithm of DGPT based on drivers' feedback, and controller design for following the preceding vehicle. 

\subsection{IRL-based offline personalized DGPT learning}
The input of the IRL algorithm is from either the demonstration trajectories when ACC is deactivated (for modeling) or the takeover trajectories while ACC is activated (for fine tuning). Similar to the existing ACC systems in the market, when the ACC is activated and the driver steps on the accelerator or brake pedal, the vehicle's controller will transfer control to the driver. This transfer of control is referred to as a takeover. As we assume that the demonstration trajectory we collected represents the optimal policy $\pi^*(s,a)$ for the equation (3), the goal of the IRL algorithm is to recover the linear coefficients $\boldsymbol{\alpha}$. The reward basis $\boldsymbol{\Phi}$ is predefined with sufficient descriptive ability in the space of $v$ and $g$. By using IRL, we can recover the reward function, as shown in Fig. \ref{fig:reward_desired_gap} (a). The detailed process for the algorithm can be found in our previous work \cite{Zhao2022ICRA}. 

\begin{figure*}[!t]
    \centering
    \includegraphics[width=\linewidth]{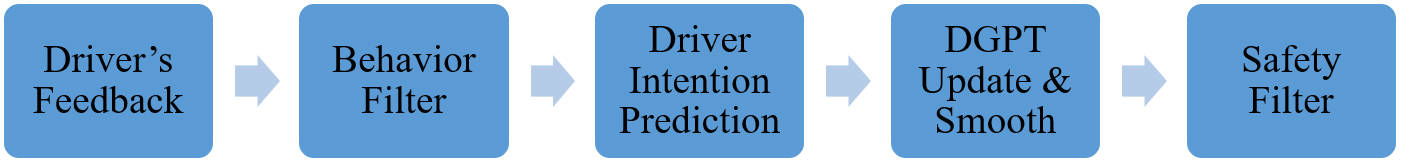}
    \caption{Flowchart of online DGPT adaptation algorithm.}
    \label{fig:adaptation}
\end{figure*}

The DGPT (i.e., the preferred gap at different speeds) is calculated using Equation (\ref{eq:4}) from the recovered reward function $r$, as shown in Fig. \ref{fig:reward_desired_gap} (b). Also, a low pass filter is applied afterward to ensure smoothness. 
\begin{equation}\label{eq:4}
    DGPT(v) = g_{\text {desired }}(v)=\arg \max _{g} r(v)
\end{equation}

\subsection{Online DGPT Adaptation}
Compared to the naturalistic car-following data that is used for training the IRL model, the online feedback data usually has a very short time period and distribute sparsely in the time domain. Therefore, additional processes are required to use this data to maintain the DGPT, as illustrated in Fig. \ref{fig:adaptation}.

First, a behavior filter is needed to ensure that only necessary updates are made to the DGPT since drivers' takeover behaviors can be very noisy and may only last for a short duration or have small inputs. Second, we need to infer the preferred steady state that the driver wishes to achieve through a short period of takeover trajectory. Intuitively, this steady state should correspond to the state when the driver stops the takeover. However, based on extensive experiments and observations, we found that this assumption is not accurate. Drivers may anticipate or prolong takeover behavior based on the speed difference with the preceding vehicle, or they may achieve a steady state through multiple takeovers. Therefore, a robust prediction mechanism is needed to determine the steady state that needs to be updated. Then, as the predicted steady states are always in scattered forms, we want to smooth updates over sufficiently large regions of the DGPT through an update and smooth module. Finally, even after the update with the aforementioned steps, the DGPT may still fall into an unreasonable range due to emergency situations or driver's mistakes, leading to potential safety hazards. Therefore, a safety space is defined for the DGPT, and a safety filter is applied to ensure that the DGPT remains bounded within this safe space. The updated DGPT then controls the ACC system if there's no drivers' takeover command.

As demonstrated, the Online DGPT Adaptation module has the potential to be a highly intricate system. In this study, we propose a simplified heuristic algorithm that adheres to the aforementioned framework in order to validate its effectiveness. The algorithm is shown in \textbf{Algorithm 1}. DGPT is the real-time personalized control reference; $v$ and $g$ are the current speed and gap, $v\_f$ is the current speed of the preceding vehicle; $p$ is the takeover status; $p\_t$ is the takeover time; $P_T$ is minimum takeover time; $K_T$ is the coast-down coefficient; $V_D$ is the maximum speed difference; $window\_size$ is the size of the smooth window; $Safe\_TG\_max$ and $Safe\_TG\_min$ are the safety time gap bounds. The safety time gap is a predefined value based on the minimum reaction time of drivers and the experience of car manufacturers.

\begin{algorithm}
\small
\SetAlgoLined
\KwData{
Input: 
$DGPT$, $v$, $v_f$, $g$, $p$, $p_t$\\
Parameters:
$P_T$, $K_T$, $V_D$, $window\_size$, $Safe\_TG\_max$, $Safe\_TG\_min$\\
Function:
moving\_average(), max(), min()
}
\KwResult{
$DGPT$
}

\For{each iteration}{
    \If{not p}{
        $update\_flag$ = $(v_f - v) \ge V_D$ or $p_t \leq P_T$\\
        \If{update\_flag}{
            \If{$v_f \ge v$}{
                $g_{desire} = g$\\
            }
            \Else{
                $g_{desire} = g(K_T \cdot (v-v\_f))$\\
            }
            $DGPT(v)=g_{desire}$\\
            moving\_average($DGPT$, $window\_size$)\\
            \For{v\_i in all possible speed}{
                $DGPT(v\_i)=max(DGPT(v\_i), Safe\_TG\_min)$\\
                $DGPT(v\_i)=min(DGPT(v\_i), Safe\_TG\_max)$\\
            }
        }
    }
}
\caption{\small{Online DGPT Adaptation}}
\end{algorithm}

\subsection{Controller Design}
In this study, we used a PID controller to control the acceleration of the ego vehicle and ensure that it follows the preceding vehicle with a desired space gap. The error between the current gap and the desired gap in the DGPT, as defined in Equation \ref{eq:5}, is used as the control input. The PID controller continuously calculates the acceleration of the vehicle based on Equation \ref{eq:6}, which takes into account the proportional, integral, and derivative components of the error. The controller aims to minimize the error and maintain a stable and safe distance between the ego vehicle and the preceding vehicle. This approach has been widely used in car-following models and has shown good performance in various scenarios.
\begin{equation}\label{eq:5}
    e(t) = DGPT(v(t)) - g(t)
\end{equation}
\begin{equation}\label{eq:6}
    a(t)=K_p\cdot e(t)+Ki\cdot\int  e(t)dt+K_p\cdot \frac{de(t)}{dt}
\end{equation}

\section{Experiments and Results} \label{sec:experiment}
 This section presents the experimental setup using the HuiL simulation platform, the metrics used to evaluate the proposed algorithm's performance, and the results and analysis of the experiments.

\subsection{Experiment Setup using HuiL Simulation}
\begin{figure}
    \centering
    \includegraphics[width=\linewidth]{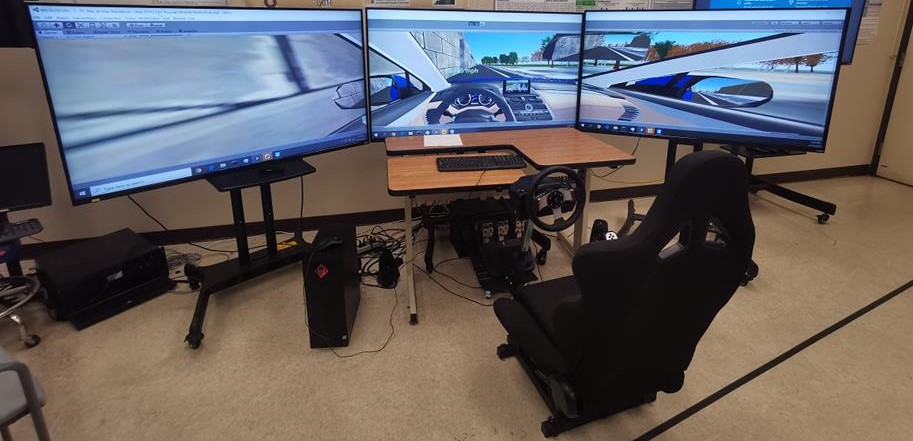}
    \caption{Human-in-the-loop simulator at the University of California, Riverside.}
    \label{fig:simulator}
\end{figure}

Validation of autonomous driving algorithms often requires consideration of safety factors, making real-world experiments difficult to conduct. Therefore, a good alternative is to use game engine-based simulations with a human-in-the-loop setup. These simulations provide testers with an immersive visual and auditory experience and can collect relatively reliable driving data. Game engines are commonly used by software developers to create video games, which typically include a physics engine, rendering engine, and scene graph for managing multiple game elements. In this study, we conducted human-in-the-loop simulations using the game engine-based driving simulator developed in our previous work \cite{wang2021digitaltwinsimulation}. The platform is built with the Unity game engine and a Logitech G27 Racing Wheel (see Fig. \ref{fig:simulator}). The simulation environment features a three-lane freeway scenario with varying weather conditions. During the test, the driver could choose to drive the ego vehicle manually or use P-ACC by monitoring the automation and pressing the accelerator or brake pedal when feeling uncomfortable. We describe the experimental setup in detail in the following subsections.

In order to test either manual driving or automatic control, the scenarios are determined by the speed profile of the leading vehicle. There are four requirements for generating the speed profile, including avoiding test driver fatigue, collecting demonstration trajectories covering a wide range of speeds for IRL model training, making the speed profile realistic, and ensuring that the scenario is unpredictable. To meet these requirements, we developed a stochastic scenario generation approach. This approach first samples short trajectory segments from naturalistic driving data, then generates a high-level random speed sequence, where each element of the sequence defines an average speed for a period of time. Next, based on the random speed sequence, the trajectory segments with corresponding average speed are randomly selected and concatenated together. Finally, a filter is applied to ensure the acceleration/deceleration of the synthesized speed profile is within the required bounds. As each driving cycle is only 300 seconds long, and a drastic change in speed may lead to abnormal driving behavior, test drivers need to take multiple driving cycles, each covering a different speed range.

We generate five different scenarios covering a range of average speeds from slow (5m/s) to fast (30m/s) using the above speed profile synthesis method. Among them, we use four speed profiles as manual driving scenarios (ACC OFF) to collect naturalistic car-following trajectories from test drivers for training the IRL-based offline personalized DGPT. A total of five test drivers participated in the experiments. A total of five test drivers participated in the experiments, each with varying driving experience ranging from 2 to 10 years, and different driving tendencies, as indicated by preliminary surveys. After obtaining the personalized DGPT for each driver, we conducted automatic control tests to validate the proposed algorithm. The driving scenarios for the automatic control tests (ACC ON) are divided into two parts: \textbf{Scenario A} involves the speed profile observed by drivers during manual driving tests, where the profile starts at a high speed of 30 m/s and gradually slows down to 5 m/s. On the other hand, \textbf{Scenario B} is a completely new scenario synthesized from naturalistic driving samples. Each scenario lasts 300 seconds.

We test four different combinations of controllers: \textbf{Predefined}, \textbf{Predefined + Online Adaptation}, \textbf{IRL}, and \textbf{IRL + Online Adaptation}. The \textbf{Predefined} ACC controller involves the driver choosing a constant time headway from high (4s), medium (3s), and low (1s) levels as the control reference, which is similar to the ACCs currently equipped on commercial vehicles. The \textbf{Predefined + Online Adaptation} setup involves the driver choosing a constant time headway as the control reference, but incorporating an online adaptation algorithm to update the control reference table based on real-time feedback, achieving a certain degree of personalization. \textbf{IRL} involves using the DGPT trained through offline IRL to control the vehicle, which has been shown in our previous studies to significantly improve drivers' comfort and trust in the system compared to ACC without personalization. Finally, \textbf{IRL + Online Adaptation} is the complete proposed framework, which uses the DGPT obtained through offline IRL as the initial control reference to control the vehicle and incorporates an online adaptation algorithm to continuously refine personalization based on real-time feedback.

\begin{table*}[]
\caption{Results of Human-in-the-Loop Simulation}
\label{tab:1}
\centering
\begin{tabular}{|ll|ll|ll|ll|ll|}
\hline
\multicolumn{1}{|l|}{\multirow{2}{*}{}}         & \multirow{2}{*}{} & \multicolumn{2}{l|}{Predefined}    & \multicolumn{2}{l|}{Predefined + Online Adaptation} & \multicolumn{2}{l|}{IRL}                   & \multicolumn{2}{l|}{IRL + Online Adaptation}                 \\ \cline{3-10} 
\multicolumn{1}{|l|}{}                          &                   & \multicolumn{1}{l|}{PoI}    & NIM  & \multicolumn{1}{l|}{PoI}             & NIM          & \multicolumn{1}{l|}{PoI}            & NIM  & \multicolumn{1}{l|}{PoI}             & NIM                   \\ \hline
\multicolumn{1}{|l|}{\multirow{2}{*}{Driver 1}} & Seen              & \multicolumn{1}{l|}{14.1\%} & 6.0  & \multicolumn{1}{l|}{\textbf{2.9\%}}  & \textbf{3.0} & \multicolumn{1}{l|}{17.6\%}         & 4.7  & \multicolumn{1}{l|}{4.7\%}           & 3.3                   \\ \cline{2-10} 
\multicolumn{1}{|l|}{}                          & Unseen            & \multicolumn{1}{l|}{13.3\%} & 7.0  & \multicolumn{1}{l|}{\textbf{7.3\%}}  & 4.7          & \multicolumn{1}{l|}{17.1\%}         & 5.0  & \multicolumn{1}{l|}{10.6\%}          & \textbf{3.0}          \\ \hline
\multicolumn{1}{|l|}{\multirow{2}{*}{Driver 2}} & Seen              & \multicolumn{1}{l|}{12.0\%} & 5.3  & \multicolumn{1}{l|}{\textbf{3.7\%}}  & 4.7          & \multicolumn{1}{l|}{4.1\%}          & 6.3  & \multicolumn{1}{l|}{9.3\%}           & \textbf{4.3}          \\ \cline{2-10} 
\multicolumn{1}{|l|}{}                          & Unseen            & \multicolumn{1}{l|}{6.7\%}  & 5.7  & \multicolumn{1}{l|}{7.8\%}           & 7.0          & \multicolumn{1}{l|}{\textbf{5.6\%}} & 5.3  & \multicolumn{1}{l|}{11.2\%}          & \textbf{4.7}          \\ \hline
\multicolumn{1}{|l|}{\multirow{2}{*}{Driver 3}} & Seen              & \multicolumn{1}{l|}{35.4\%} & 29.3 & \multicolumn{1}{l|}{10.9\%}          & 15.7         & \multicolumn{1}{l|}{31.4\%}         & 11.0 & \multicolumn{1}{l|}{\textbf{3.2\%}}  & \textbf{3.0}          \\ \cline{2-10} 
\multicolumn{1}{|l|}{}                          & Unseen            & \multicolumn{1}{l|}{17.0\%} & 12.7 & \multicolumn{1}{l|}{\textbf{3.1\%}}  & \textbf{6.7} & \multicolumn{1}{l|}{25.9\%}         & 13.7 & \multicolumn{1}{l|}{11.3\%}          & 9.0                   \\ \hline
\multicolumn{1}{|l|}{\multirow{2}{*}{Driver 4}} & Seen              & \multicolumn{1}{l|}{26.0\%} & 5.7  & \multicolumn{1}{l|}{10.4\%}          & 4.0          & \multicolumn{1}{l|}{3.6\%}          & 1.7  & \multicolumn{1}{l|}{2.6\%}           & \textbf{1.3}          \\ \cline{2-10} 
\multicolumn{1}{|l|}{}                          & Unseen            & \multicolumn{1}{l|}{21.9\%} & 5.3  & \multicolumn{1}{l|}{10.0\%}          & 3.0          & \multicolumn{1}{l|}{8.7\%}          & 2.7  & \multicolumn{1}{l|}{2.4\%}           & \textbf{1.3}          \\ \hline
\multicolumn{1}{|l|}{\multirow{2}{*}{Driver 5}} & Seen              & \multicolumn{1}{l|}{29.8\%} & 16.0 & \multicolumn{1}{l|}{13.8\%}          & 10.3         & \multicolumn{1}{l|}{16.3\%}         & 7.3  & \multicolumn{1}{l|}{\textbf{10.8\%}} & \textit{\textbf{6.3}} \\ \cline{2-10} 
\multicolumn{1}{|l|}{}                          & Unseen            & \multicolumn{1}{l|}{26.6\%} & 14.3 & \multicolumn{1}{l|}{11.6\%}          & 4.3          & \multicolumn{1}{l|}{14.9\%}         & 6.0  & \multicolumn{1}{l|}{\textbf{9.3\%}}  & \textbf{4.3}          \\ \hline
\multicolumn{2}{|l|}{Average}                                       & \multicolumn{1}{l|}{20.3\%} & 10.7 & \multicolumn{1}{l|}{8.2\%}           & 6.3          & \multicolumn{1}{l|}{14.5\%}         & 6.4  & \multicolumn{1}{l|}{7.5\%}           & 4.1                   \\ \hline
\end{tabular}
\end{table*}

\subsection{Results and Analysis}
In this section, we present the results of the experiments and analyze the data. We use Percentage of Interruption (PoI) and Number of Interruption-per-Minute (NIM) to quantitatively measure the driver's comfort and trust in the P-ACC system. PoI denotes the time percentage when the driver steps onto the acceleration pedal or brake pedal, and NIM denotes the number of times the driver steps onto the pedals. The best performance is marked in bold for each scenario. As shown in TABLE \ref{tab:1}, both the PoI and NIM have been greatly reduced when the complete proposed system is applied and compared with the baseline controller. Experiments show that the average PoI reduction was 62.8\% and the average NIM decreased by 62.2\% compared to the predefined ACC settings. In some cases, the PoI and NIM decreased up to 91.5\% and 82.2\%, respectively. This indicates the drivers are more satisfied with automatic car-following based on the proposed P-ACC. This advantage is observed in both seen and unseen driving scenarios, indicating the robustness of both offline and online modules in adapting to new situations. Notably, the \textbf{IRL + Online Adaptation} controller outperforms other automation control methods in the NIM metric significantly. However, in some cases, the \textbf{Predefined + Online Adaptation} controller performs better. This indicates that online adaptation can quickly adapt to the drivers' preferences when the selected default settings match their driving style. Comparing the \textbf{Predefined + Online Adaptation} controller and the \textbf{IRL} controller, the \textbf{Predefined + Online Adaptation} controller performs better in most cases. This suggests that online adaptation is better suited to real-time driving preferences than offline learning. Particularly, offline learning may fail when there are significant changes in driving scenarios or driver mood. In such cases, online adaptation is essential.

\section{Conclusion and Future Work} \label{sec:conclusion}
In summary, vehicle automation and Advanced Driver Assistance Systems (ADAS) are playing an increasingly important role in enhancing driving safety and comfort. However, pre-defined settings may not always align with individual driver preferences and styles. The emergence of personalized ADAS (P-ACC) aims to solve this problem. While previous research has primarily focused on using historical driving data to create personalized controllers, this study proposes a novel cloud-vehicle collaborative P-ACC framework that includes both offline and online components. By recording the driver's naturalistic car-following trajectories and utilizing IRL in the cloud to train the personalized model (e.g., DGPT), the offline component can obtain the driver's preference before the trip. Then, while the ACC is activated en-route, the online component updates the corresponding DGPT in real time by adapting to the driver's feedback (i.e., takeover of the control). Additionally, with the help of incremental learning achieved through retraining the model based on driver's takeover trajectories, the model gradually becomes more consistent with the driver's driving preferences. Human-in-the-loop (HuiL) simulation experiments demonstrate that this method can significantly reduce driver interventions in the automatic control system, where average PoI has decreased by up to 62.8\%, and average NIM has decreased by up to 62.2\% for each scenario. This personalized approach can help to ensure a more comfortable experience while also increasing driver trust and usage of this type of ADAS.

In the future, we plan to develop more sophisticated online adaptation methods that can better estimate whether the driver has reached a satisfactory state and can more accurately estimate the preferred state's value. Additionally, improving the global update of the entire Driving Gap Preference Table (DGPT) based on these discrete values will help further gear the system towards each driver's needs. Furthermore, conducting real-world tests with actual drivers would provide valuable insights into the effectiveness of the proposed framework. Finally, incorporating additional sensor data and considering more complex driving scenarios, such as intersections and merging lanes, will help enhance the performance and adaptability of the P-ACC system.

\section*{Acknowledgment}
This work is sponsored by the ``Digital Twin'' project of InfoTech Labs, Toyota Motor North America R\&D. The authors would like to sincerely thank, Yejia Liao, and Ahmadreza Moradipari for their participation in the human-in-the-loop experiments.

The contents of this work only reflect the views of the authors, who are responsible for the facts and the accuracy of the data presented herein. The contents do not necessarily reflect the official views of Toyota Motor North America.

\bibliographystyle{IEEEtran}
\bibliography{refs}
\end{document}